\pdfoutput=1
\documentclass[a4paper,10pt]{article}

\usepackage[T1]{fontenc}
\usepackage{lmodern}
\usepackage{microtype}
\usepackage[utf8]{inputenc}
\usepackage[USenglish,UKenglish]{babel}
\usepackage{fullpage}
\usepackage{graphicx}
\usepackage{amsmath,amssymb,amsthm}
\usepackage{authblk}
\usepackage[noadjust]{cite}
\usepackage{url}
\usepackage[hidelinks]{hyperref}
\hypersetup{pdftitle={Exact and Approximation Algorithms for Computing Smallest SESs and BMSs},pdfauthor={Hiroki Shibata and Hideo Bannai}}
\usepackage{thmtools,thm-restate}
\usepackage[capitalise,noabbrev]{cleveref}

\declaretheorem[name=Theorem]{theorem}
\declaretheorem[name=Lemma,sibling=theorem]{lemma}
\declaretheorem[name=Corollary,sibling=theorem]{corollary}
\declaretheorem[name=Example,sibling=theorem,style=definition]{example}

\title{Exact and Approximation Algorithms for Computing Smallest SESs and BMSs}
\author[1]{Hiroki Shibata\thanks{\texttt{shibata.hiroki.753@s.kyushu-u.ac.jp}}}
\author[2]{Hideo Bannai\thanks{\texttt{bannai.h.557a@m.isct.ac.jp}}}
\affil[1]{Joint Graduate School of Mathematics for Innovation, Kyushu University, Japan}
\affil[2]{M\&D Data Science Center, Institute of Biomedical Engineering, Institute of Science Tokyo, Japan}
\date{{\selectlanguage{USenglish}\today}}

\newcommand{\calT}{\mathcal{T}}
\newcommand{\sgn}{\operatorname{sgn}}
\DeclareMathOperator*{\argmax}{arg\,max}
\newcommand{\Set}{\mathsf{Set}}
\newcommand{\Eq}{\mathsf{Eq}}
\newcommand{\Ch}{\mathsf{Ch}}
\newcommand{\SES}[1]{\mathsf{SES}(#1)}
\begin{document}

\maketitle

\begin{abstract}
	Compression schemes represent repetitive strings compactly by using repeated substrings.
	Computing the minimum size of a representation of a given input string is important for compression and for understanding relationships between compression schemes, although this task is NP-hard for several schemes.
In this paper, we introduce exact and approximation algorithms for finding smallest substring equation systems (SESs) and bidirectional macro schemes (BMSs).
	Specifically, for an input string of length $n$, we give deterministic and randomized algorithms for computing a smallest SES using $O^*(4^n)$ and $O^*(2^n)$ time and space, respectively.
	We also give a deterministic algorithm for computing a smallest BMS using $O^*(2^n)$ time and space.
	We further give deterministic and randomized $(1+\alpha)$-approximation algorithms for smallest SESs using $O^*(4^{e^{-\alpha}n})$ and $O^*(2^{e^{-\alpha}n})$ time and space, respectively, for every $\alpha \ge 0$.
	A conversion from SESs to BMSs gives $4(1+\alpha)$-approximation algorithms for smallest BMSs with the same time and space bounds.
	All time bounds hold in the worst case, and the randomized algorithms return correct answers with high probability.
Our algorithms adapt several approaches to exact and approximation algorithms for combinatorial problems, such as dynamic programming over determinant expansions and a greedy algorithm for monotone submodular functions.
\end{abstract}
\section{Introduction}
\label{sec:introduction}

Dictionary compression methods such as Lempel--Ziv parsing~\cite{ZivL77} efficiently compress strings by representing equal substrings through references.
The minimum size of such a representation gives a repetitiveness measure of the string.
These measures help compare the power of compression methods and serve as parameters for compressed data structures and algorithms~\cite{Navarro22}.

Bidirectional macro schemes (BMSs)~\cite{StorerS82} and substring equation systems (SESs)~\cite{ShibataB26,GawrychowskiKRRW20} are representative compression schemes that use substring equalities and whose minimum sizes are studied as repetitiveness measures.
A BMS partitions a string $w$ into phrases and represents each phrase by an explicit symbol or a reference to an equal substring.
An SES specifies substring equalities and symbol assignments that uniquely determine the string without requiring a partition into phrases, and its minimum size has recently been studied as a repetitiveness measure~\cite{ShibataB26}.

Efficient algorithms for finding smallest representations in these compression schemes are useful both for compression and for studying size bounds between repetitiveness measures.
However, computing many repetitiveness measures, including the smallest grammar, the smallest BMS, and the smallest SES, is NP-hard~\cite{CharikarLLPPSS05,StorerS82,ShibataB26}.
These hardness results motivate exact algorithms with exponential running time and approximation algorithms with guarantees on the output size.

For these NP-hard problems, few efficient exact algorithms with worst-case running-time guarantees are known.
For the smallest grammar problem, an $O^*(3^n)$-time algorithm is known~\cite{CaselFGGS21} and a recent result shows an $O^*(1.8723^n)$-time algorithm~\cite{Chen26}.
In contrast, such single-exponential-time exact algorithms have not been given for BMSs or SESs.
Several methods based on MAX-SAT solvers or answer set programming have been proposed to compute these measures exactly in practice~\cite{BannaiGIKKN22,KawamotoIKB24,BanbaraBHKMN26}, but they provide no worst-case time guarantees.

Although the difficulty of exact computation makes approximation algorithms a natural alternative, no polynomial-time algorithm is known to approximate any of these measures within a constant factor.
Moreover, it is known that approximating smallest grammars or SESs within a factor of $1+\varepsilon$ is NP-hard for some constant $\varepsilon > 0$~\cite{CharikarLLPPSS05,ShibataB26}.
These limitations motivate exponential-time approximation algorithms with better approximation ratios than polynomial-time methods and faster running times than exact algorithms, as studied for classical combinatorial optimization problems including weighted MAX-2-SAT~\cite{Williams05} and weighted set cover~\cite{CyganKW09}.
Such approximation algorithms could make near-optimal compressed representations more practical to compute and help study the relationships between repetitiveness measures.

\subsection*{Our Contributions}

In this paper, we give exact and approximation algorithms for computing smallest SESs and BMSs.
Let $s(w)$ and $b(w)$ denote the sizes of a smallest SES and a smallest BMS for a string $w$, respectively.
Our main results are stated in the following theorems.

\begin{restatable}[Deterministic Algorithm for Smallest SESs]{theorem}{sesresult}
\label{thm:ses-exact-det}
Given an input string $w$ of length $n$, the value $s(w)$ and a smallest SES representing $w$ can be computed deterministically in $O^*(4^n)$ time and space.
\end{restatable}

\begin{restatable}[Randomized Algorithm for Smallest SESs]{theorem}{sesrandresult}
\label{thm:ses-exact-rand}
Given an input string $w$ of length $n$, the value $s(w)$ and a smallest SES representing $w$ can be computed by a randomized algorithm in $O^*(2^n)$ worst-case time and space with probability at least $1 - 2^{-n}$.
\end{restatable}

\begin{restatable}[Approximation Algorithms for Smallest SESs]{theorem}{sesapproxresult}
\label{thm:ses-approximation}
For every $\alpha \ge 0$, an SES representing $w$ of size at most $(\alpha + 1)s(w)$ can be computed deterministically in $O^*(4^{e^{-\alpha} n})$ time and space, or with probability at least $1 - 2^{-n}$ in $O^*(2^{e^{-\alpha} n})$ worst-case time and space.
\end{restatable}

\begin{restatable}[Deterministic Algorithm for Smallest BMSs]{theorem}{bmsresult}
	\label{thm:bms}
	Given an input string $w$ of length $n$, the value $b(w)$ and a smallest BMS representing $w$ can be computed deterministically in $O^*(2^n)$ time and space.
\end{restatable}

Combining the conversion from SESs to BMSs~\cite{ShibataB26} with \cref{thm:ses-approximation} also gives a $4(\alpha+1)$-approximation for smallest BMSs with the same time and space bounds.

\subsection*{Overview of Our Methods}
As an introduction to the technical details, we give a brief overview of our algorithms and their connections to existing work.

For the exact algorithms for the smallest SES problem, we first reduce the problem to finding a spanning tree minimizing the number of groups of edges corresponding to constraints.
Following an existing approach for graph connectivity problems~\cite{BodlaenderCKN15}, we use the Matrix--Tree theorem and dynamic programming over determinant expansions of a reduced incidence matrix to find such a spanning tree.
For the randomized algorithm, we reduce the number of states in dynamic programming by using random variables.

Our approximation algorithms combine greedy selection for a monotone submodular function with the exact algorithm to obtain a trade-off between the approximation ratio and the running time.
At the first step, we select constraints greedily to reduce the number of connected components and stop at a threshold controlled by $\alpha$.
Next, we use the exact algorithms to find a smallest SES containing the selected constraints.
This combination of greedy selection and exact computation has the same structure and gives a similar trade-off as the algorithms for the weighted set cover problem~\cite{CyganKW09}.

For BMSs, the fact that phrases partition the string allows us to construct a faster deterministic exact algorithm.
 \section{Preliminaries}
\label{sec:preliminaries}

\subparagraph*{Strings.}

For an integer $n \ge 0$, let $[n] = \{1, \dots, n\}$.
For integers $a \le b$, let $[a,b] = \{a, \dots, b\}$.
Let $\Sigma$ be an alphabet with $|\Sigma| \geq 2$.
A string $w$ of length $|w| = n$ is a sequence $w[1]\cdots w[n]$ of symbols of $\Sigma$.
For $1 \le i \le j \le n$, we write $w[i..j] = w[i]\cdots w[j]$ for the substring of $w$ from position $i$ to position $j$.
For $j < i$, we define $w[i..j]$ to be the empty string.
Throughout the paper, we denote a nonempty input string by $w$ and its length by $n$.
We write $O^*(f(n))$ for $O(f(n)\cdot n^{O(1)})$, i.e., ignoring polynomial factors in $n$.

\subparagraph*{Substring equation systems.}

A \emph{substring equation system} (SES) for a string of length $n$ is a triple $(n,\Eq,\Ch)$, where
$\Eq$ is a finite set of \emph{substring equations} $(\{i,j\},\ell)$ with $1\le i<j\le n$ and $1\le\ell\le n-j+1$, representing the constraint $w[i..i+\ell-1]=w[j..j+\ell-1]$,
and $\Ch$ is a finite set of \emph{symbol assignments} $(k,c)$ with $k\in[n]$ and $c\in\Sigma$, representing the constraint $w[k]=c$.
The system \emph{represents} $w$ if $w$ is the unique string in $\Sigma^n$ satisfying all constraints.
Its \emph{size} is $|\Eq|+|\Ch|$, and $s(w)$ is the minimum size of an SES representing $w$.
Given a string $w$ of length $n$ and an SES $(n, \Eq, \Ch)$, we can decide whether $w$ is the unique string satisfying all constraints in $O(n+|\Eq|+|\Ch|)$ time~\cite{GawrychowskiKRRW20}.

\subparagraph*{Bidirectional Macro Schemes.}
A \emph{bidirectional macro scheme} (BMS)~\cite{StorerS82} for $w$ partitions $w$ into consecutive \emph{phrases} $w=f_1\cdots f_k$.
Each phrase $f=w[i..i+\ell-1]$ is either a single symbol stored explicitly ($\ell=1$) or a \emph{copy}, given by a source position $p\ne i$ with $w[i..i+\ell-1]=w[p..p+\ell-1]$.
The scheme induces a \emph{reference function} $\varphi\colon[n]\to\{0\}\cup[n]$ with $\varphi(i)=0$ if $w[i]$ is stored explicitly and $\varphi(i+h)=p+h$ for $0\le h<\ell$ if $w[i..i+\ell-1]$ is a copy with source $p$.
The reference function is \emph{valid} if for every $i\in[n]$ some iterate $\varphi^k(i)$ equals $0$, i.e., the directed graph with arcs $i\to\varphi(i)$ has no cycle.
If the reference function is valid, following references recovers every symbol from an explicitly stored symbol and hence determines $w$ uniquely; we call such a scheme \emph{valid}.
Let $b(w)$ be the minimum number of phrases of a valid BMS for $w$.

 \section{Exact Algorithms for Computing a Smallest SES}
\label{sec:exact_ses}

In this section, we give deterministic and randomized algorithms for computing a smallest SES.

\sesresult*

\sesrandresult*

\subsection{Characterization of the Minimum SES} \label{sse:char-minSES}

\begin{figure}[t]
	\centering
	\includegraphics[width=\textwidth]{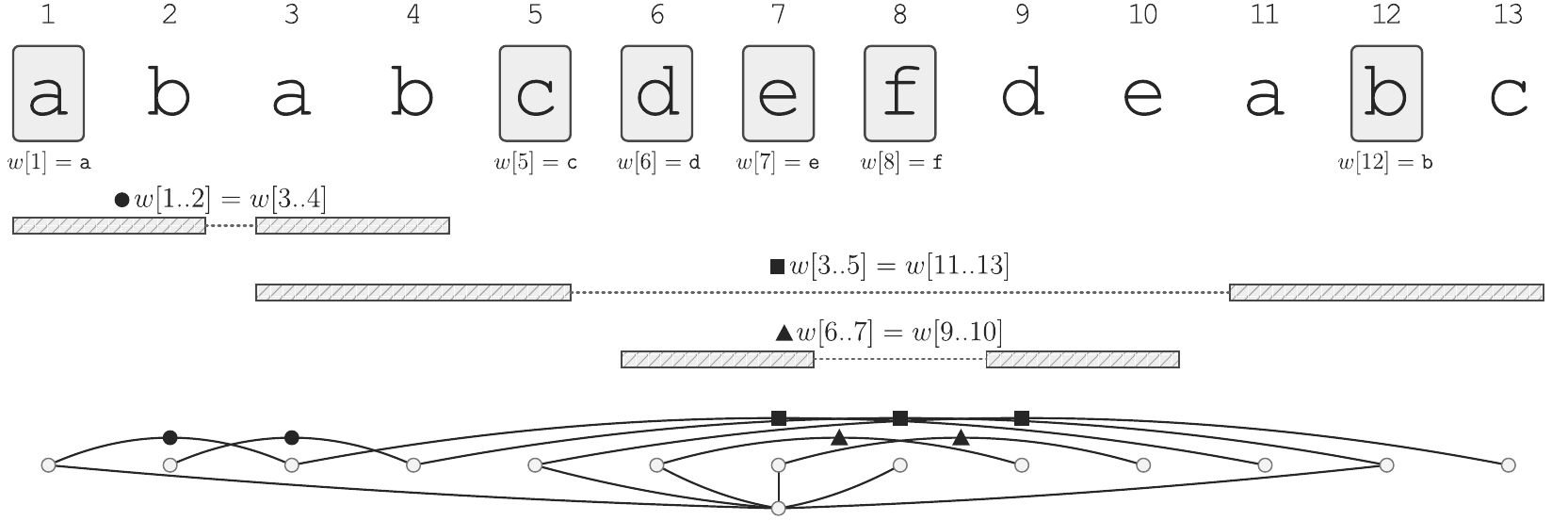}
	\caption{
		An SES for $w = \mathtt{ababcdefdeabc}$ (top) and the graph formed by its constraints (bottom).
		Rounded boxes indicate symbol assignments in $\Ch$, and each pair of hatched rectangles joined by a dotted line represents a substring equation in $\Eq$.
		Each edge corresponding to a constraint of the SES is shown in black.
		Edges that connect string positions and belong to the same run are drawn with the same mark.
}
	\label{fig:ababcdefdeabc-ses-graph}
\end{figure}

Let $G(w)$ be the graph on $\{0\} \cup [n]$ with edge set $E(w) = \{\{p, q\} \mid 1 \le p < q \le n,\ w[p] = w[q]\} \cup \{\{0, v\} \mid v \in [n]\}$.
A \emph{run} of $E(w)$ is either a singleton $\{\{0, v\}\} \subseteq E(w)$ or a maximal subset of $E(w)$ of the form $\{\{p, q\}, \{p+1, q+1\}, \dots, \{p+\ell-1, q+\ell-1\}\}$ for integers $1 \le p < q$ and $\ell \ge 1$.
The run $\{\{0, v\}\}$ represents the assignment $(v, w[v])$, whereas a run $\{\{p+h, q+h\} \mid h \in [0, \ell-1]\}$ with $p \ge 1$ corresponds to the set of edges induced by the maximal substring equation $(\{p, q\}, \ell)$.
Let $\mathcal{R}(w)$ be the set of runs of $E(w)$, which forms a partition of $E(w)$.
For $\mathcal Q \subseteq \mathcal R(w)$, let $E(\mathcal Q) = \bigcup_{R \in \mathcal Q}R$.
For $\mathcal Q \subseteq \mathcal R(w)$, let $\SES{\mathcal Q} = (n, \Eq, \Ch)$ be the SES with size $|\mathcal Q|$ whose substring equations and symbol assignments are represented by the runs in $\mathcal Q$.
\Cref{fig:ababcdefdeabc-ses-graph} illustrates an SES and the graph formed by its constraints.

The next lemma characterizes SESs representing $w$ by the connectivity of the corresponding graph.
\begin{lemma}
	\label{lem:ses-connected}
	For $\mathcal Q \subseteq \mathcal R(w)$, $\SES{\mathcal Q}$ represents $w$ if and only if the graph $(\{0\} \cup [n], E(\mathcal Q))$ is connected.
\end{lemma}
\begin{proof}
	All constraints of $\SES{\mathcal Q}$ are satisfied by $w$.
	If the graph is connected, every position is connected by edges of substring equations to a position with a symbol assignment.
	These constraints fix the symbol at every position, so $\SES{\mathcal Q}$ represents $w$.
	If the graph is disconnected, a connected component not containing $0$ has no symbol assignment and no substring equation relates a position in this component to a position outside it.
	Any symbol can be assigned to all positions in this component, so the solution is not unique and $\SES{\mathcal Q}$ does not represent $w$.
\end{proof}

For $T \subseteq E(w)$, let $\mathcal Q(T)$ be the set of runs in $\mathcal R(w)$ containing at least one edge of $T$.
The following lemma reduces computing a smallest SES to finding a spanning tree that minimizes the number of runs containing its edges.
\begin{lemma}
	\label{lem:ses-char}
	$s(w) = \min\{|\mathcal Q(T)| \mid T \text{ is a spanning tree of }G(w)\}$.
\end{lemma}
\begin{proof}
	($\le$) Let $T$ be a spanning tree of $G(w)$.
	Since $T \subseteq E(\mathcal Q(T))$, the graph $(\{0\} \cup [n], E(\mathcal Q(T)))$ is connected.
	By \cref{lem:ses-connected}, $\SES{\mathcal Q(T)}$ represents $w$ and has size $|\mathcal Q(T)|$.
	Thus $s(w) \le |\mathcal Q(T)|$.

	($\ge$) Let $(n, \Eq, \Ch)$ be a smallest SES representing $w$.
	We extend each substring equation in both directions to a maximal substring equation satisfied by $w$, so that its induced edges form a run in $\mathcal R(w)$.
	Let $\mathcal Q$ consist of these runs and the singleton runs for the symbol assignments in $\Ch$.
	Every solution of the extended constraints satisfies the original constraints and $w$ satisfies the extended constraints, so $\SES{\mathcal Q}$ also represents $w$.
	By \cref{lem:ses-connected}, the graph $(\{0\} \cup [n], E(\mathcal Q))$ is connected and contains a spanning tree $T$ of $G(w)$.
	Since $\mathcal Q(T) \subseteq \mathcal Q$, we have $|\mathcal Q(T)| \le |\mathcal Q| \le |\Eq| + |\Ch| = s(w)$.
\end{proof}

\subsection{A Deterministic Algorithm}
\label{sec:ses-deterministic}

\begin{figure}[t]
	\centering
	\includegraphics[width=\textwidth]{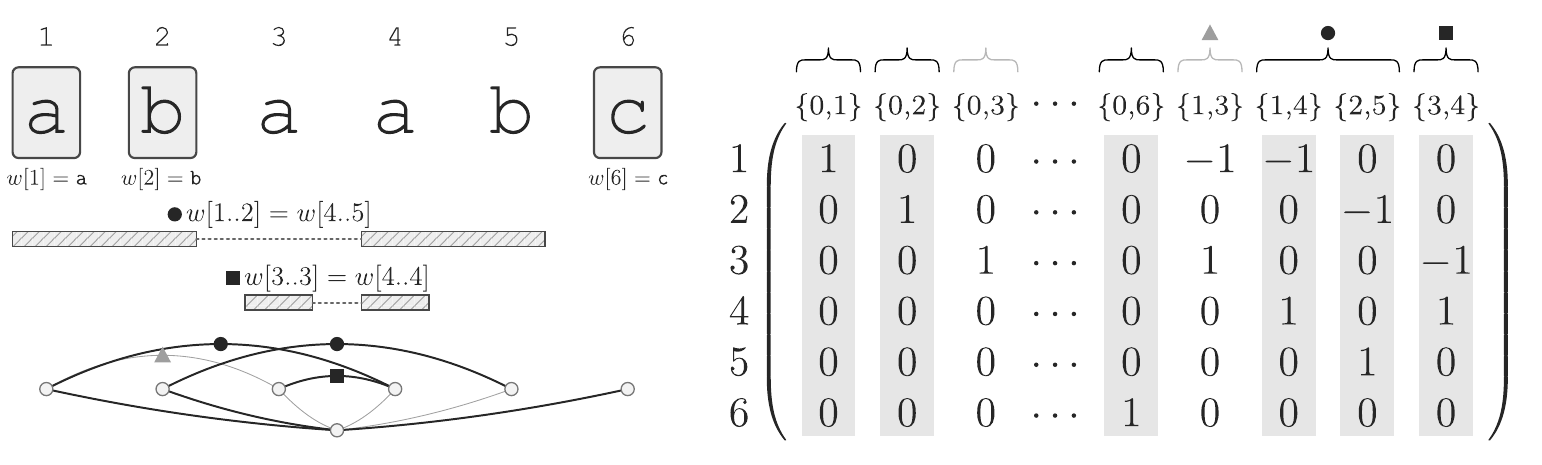}
	\caption{
		An SES for $w = \mathtt{abaabc}$ (upper left), the graph formed by its constraints (lower left), and the corresponding matrix $B$ (right).
		The drawing conventions are the same as in \cref{fig:ababcdefdeabc-ses-graph}, with unselected edges of $G(w)$ also shown in light gray.
		In the right figure, each row corresponds to a string position, each column corresponds to an edge of $G(w)$, and each brace represents a run.
		Columns corresponding to selected edges are shaded gray, and the runs containing these edges are highlighted by black braces.
}
	\label{fig:abaabc-ses-matrix}
\end{figure}

We give an efficient algorithm for computing $s(w)$ based on the characterization in \cref{lem:ses-char}.
We write $E(w) = \{e_1, \dots, e_m\}$ with $m = |E(w)|$ so that the edges belonging to the same run have consecutive indices.
For $1 \le j < i \le m$, let $c(j, i) = 1$ if $e_j$ and $e_i$ belong to different runs and let $c(j, i) = 0$ otherwise.
We define $c(0, i) = 1$ for $i \in [m]$.
For $T = \{e_{j_1}, \dots, e_{j_{|T|}}\}$ with $0 = j_0 < j_1 < \dots < j_{|T|}$, we have $|\mathcal Q(T)| = \sum_{r=1}^{|T|}c(j_{r-1}, j_r)$.

We define a matrix $B \in \mathbb Z^{[n] \times E(w)}$ by
\begin{equation}
	\label{eq:ses-incidence}
	B[u, e] = \begin{cases}
		-1 & \text{if } u = i, \\
		1  & \text{if } u = j, \\
		0  & \text{otherwise},
	\end{cases}
\end{equation}
where $u \in [n]$ and $e = \{i, j\} \in E(w)$ with $i < j$.
For $T \subseteq E(w)$ with $|T| = n$, let $B_T$ be the $n \times n$ matrix obtained from $B$ by taking the columns corresponding to the edges in $T$ while preserving their order.
\Cref{fig:abaabc-ses-matrix} illustrates the correspondence between the graph and the matrix $B$.

The matrix $B_T$ is essentially the incidence matrix of the graph $(\{0\} \cup [n], T)$ by deleting the row indexed by $0$.
Thus, we can use the existing characterization of spanning trees by determinants~\cite[Lemma~4.1]{BodlaenderCKN15} and obtain the following result.
\begin{lemma}[{\cite[Lemma~4.1]{BodlaenderCKN15}}]
	\label{lem:tree-det}
	For every $T \subseteq E(w)$ with $|T| = n$, $|\det B_T| = 1$ if $T$ is a spanning tree of $G(w)$ and $|\det B_T| = 0$ otherwise.
\end{lemma}

For $k \in [0, n]$, let $\calT(w, k) = \{T \subseteq E(w) \mid |T| = n,\ |\mathcal Q(T)| = k\}$.
By \cref{lem:tree-det}, the number of spanning trees $T$ with $|\mathcal Q(T)| = k$ is
\begin{equation}
	\label{eq:ses-count}
	N_k = \sum_{T \in \calT(w, k)} (\det B_T)^2.
\end{equation}
Combining the above formula with \cref{lem:ses-char}, we obtain $s(w) = \min\{k \in [0, n] \mid N_k > 0\}$.
In the following, we show how to compute all values $N_k$.

For $S \subseteq [n]$ and $j \in [n]$, let ${\mathrm{rank}}(S, j) = |\{s \in S \mid s > j\}|$.
For a sequence of integers $(j_1, \dots, j_t)$, its \emph{inversion number} $\mathrm{inv}(j_1, \dots, j_t) = \sum_{i=1}^{t} {\mathrm{rank}}(\{j_1, \dots, j_{i-1}\}, j_i)$ is the number of pairs of indices $u < v$ with $j_u > j_v$.
For a positive integer $m$ and $0 \le t \le m$, let $P_{m,t}$ be the set of sequences of $t$ distinct elements of $[m]$, and $P_m = P_{m, m}$ be the set of permutations of $[m]$.
For $\pi \in P_{m,t}$, we define its sign by $\sgn(\pi) = (-1)^{\mathrm{inv}(\pi(1), \dots, \pi(t))}$.
For the empty sequence $\pi$, we have $\mathrm{inv}(\pi) = 0$ and $\sgn(\pi) = 1$.

For an edge set $T = \{e_{j_1}, \dots, e_{j_n}\}$ with $j_1 < \dots < j_n$, the $i$-th column of $B_T$ corresponds to the column of $B$ indexed by $e_{j_i}$.
Hence, the Leibniz formula gives
\[
	\det B_T = \sum_{\pi \in P_n} \sgn(\pi)\prod_{i=1}^{n} B[\pi(i), e_{j_i}],
\]
and
\[
	\begin{aligned}
		(\det B_T)^2
		 & = \left(\sum_{\pi_1 \in P_n} \sgn(\pi_1)\prod_{i=1}^{n} B[\pi_1(i), e_{j_i}]\right)
		\left(\sum_{\pi_2 \in P_n} \sgn(\pi_2)\prod_{i=1}^{n} B[\pi_2(i), e_{j_i}]\right)      \\
		 & = \sum_{\pi_1, \pi_2 \in P_n}
		\left(\sgn(\pi_1)\prod_{i=1}^{n} B[\pi_1(i), e_{j_i}]\right)
		\left(\sgn(\pi_2)\prod_{i=1}^{n} B[\pi_2(i), e_{j_i}]\right).
	\end{aligned}
\]

Let $T = \{e_{j_1}, \dots, e_{j_\ell}\}$ with $j_1 < \dots < j_\ell$ and $0 \le \ell \le n$.
For $\pi_1, \pi_2 \in P_{n,\ell}$, we define $v(T, \pi_1, \pi_2)$ as
\[
	v(T, \pi_1, \pi_2)
	= \left(\sgn(\pi_1)\prod_{i=1}^{\ell} B[\pi_1(i), e_{j_i}]\right)
	\left(\sgn(\pi_2)\prod_{i=1}^{\ell} B[\pi_2(i), e_{j_i}]\right).
\]
When $T$ and both sequences are empty, $v(T, \pi_1, \pi_2) = 1$.
By substituting this expansion into \eqref{eq:ses-count}, we obtain
\begin{equation}
	\label{eq:ses-expansion}
	N_k = \sum_{T \in \calT(w, k)} (\det B_T)^2
	= \sum_{T \in \calT(w, k)} \sum_{\pi_1, \pi_2 \in P_n} v(T, \pi_1, \pi_2).
\end{equation}

For $T \subseteq E(w)$ with $|T| = t$ where $1 \le t \le n$ and $\pi_1, \pi_2 \in P_{n,t}$,
we consider computing the values $v(T, \pi_1, \pi_2)$ and $|\mathcal Q(T)|$ from corresponding values for an edge set obtained by removing one edge from $T$.
Let $e_i$ be the last edge of $T$ in the order $e_1, \dots, e_m$ and let $T' = T \setminus \{e_i\}$.
Let $\pi'_1, \pi'_2 \in P_{n,t-1}$ be obtained by deleting the last values $u_1, u_2$ from $\pi_1, \pi_2$, respectively.
Let $S'_1 = \Set(\pi'_1)$ and $S'_2 = \Set(\pi'_2)$, where $\Set(A)$ denotes the set of values of a sequence $A$.
For $a \in \{1, 2\}$, we have $\mathrm{inv}(\pi_a) - \mathrm{inv}(\pi'_a) = \mathrm{rank}(S'_a, u_a)$.
By this identity and the definition of $v$, we have

\begin{equation}
	\label{eq:ses-prefix}
	\begin{aligned}
		v(T, \pi_1, \pi_2)
		 & = v(T', \pi'_1, \pi'_2)                                                                        \\
		 & \quad\cdot (-1)^{\mathrm{rank}(S'_1, u_1) + \mathrm{rank}(S'_2, u_2)} B[u_1, e_i] B[u_2, e_i].
	\end{aligned}
\end{equation}
By the definition of $c$, we have $|\mathcal Q(T)| = |\mathcal Q(T')| + c(j, i)$, where $j$ is the index of the last edge of $T'$ with $j = 0$ if $T'$ is empty.

We identify the information needed to update $v(T, \pi_1, \pi_2)$ and $|\mathcal Q(T)|$ when adding an edge to $T$ and appending a row index to each sequence.
For $\pi_1, \pi_2 \in P_{n,|T|}$, let $S_1 = \Set(\pi_1)$ and $S_2 = \Set(\pi_2)$.
By \eqref{eq:ses-prefix}, we can compute the value of $v$ after adding an edge and appending a row index to each sequence from $v(T, \pi_1, \pi_2)$ and the sets $S_1, S_2$.
To update $|\mathcal Q(T)|$, we also need $k = |\mathcal Q(T)|$ and the position $i$ of the last edge of $T$.

Based on the above observation, we use dynamic programming to group triples that have the same values of $i$, $S_1$, $S_2$, and $k$.
For $S_1, S_2 \subseteq [n]$ and $k \in [0, n]$, let $\mathcal D(i, S_1, S_2, k)$ be the set of triples $(T, \pi_1, \pi_2)$ with $T \subseteq E(w)$, $i = \max\{j \mid e_j \in T\}$, $|\mathcal Q(T)| = k$, and $\pi_1, \pi_2 \in P_{n,|T|}$ satisfying $\Set(\pi_1) = S_1$ and $\Set(\pi_2) = S_2$.
For $i = 0$, $\mathcal D(0, \emptyset, \emptyset, 0)$ consists of the triple with $T, \pi_1, \pi_2$ all empty, and all other sets $\mathcal D(0, S_1, S_2, k)$ are empty.
We define
\[
	D[i, S_1, S_2, k]
	= \sum_{(T, \pi_1, \pi_2) \in \mathcal D(i, S_1, S_2, k)} v(T, \pi_1, \pi_2).
\]

The initial values are $D[0, \emptyset, \emptyset, 0] = 1$ and $0$ for all other states with $i = 0$.
We also set $D[i, S_1, S_2, k] = 0$ if $k \notin [0, n]$ or $|S_1| \ne |S_2|$.
For $i \in [1, m]$, we can compute the value of $D[i, S_1, S_2, k]$ by

\begin{equation}
	\label{eq:ses-dp}
	\begin{aligned}
		D[i, S_1, S_2, k]
		 & = \sum_{j=0}^{i-1} \sum_{\substack{u \in S_1 \\ v \in S_2}}
		                      (-1)^{\mathrm{rank}(S_1 \setminus \{u\}, u) + \mathrm{rank}(S_2 \setminus \{v\}, v)} B[u, e_i] B[v, e_i] \\
		 & \qquad\cdot D[j, S_1 \setminus \{u\}, S_2 \setminus \{v\}, k-c(j, i)].
	\end{aligned}
\end{equation}
Here the index $j$ corresponds to the index of the last edge after removing $e_i$ from $T$, with $j = 0$ for the empty edge set.

\begin{restatable}{lemma}{sesdpcorrect}
	\label{lem:ses-dp-correct}
	The initial values and the recurrence in \eqref{eq:ses-dp} correctly compute $D[i, S_1, S_2, k]$.
\end{restatable}
\begin{proof}
	We prove the claim by induction on $i$.
	For $i = 0$, the only triple consists of the empty edge set and two empty sequences, and its value is $1$.

	For $i \ge 1$, let $(T, \pi_1, \pi_2) \in \mathcal D(i, S_1, S_2, k)$ and let $T' = T \setminus \{e_i\}$.
	Let $u, v$ be the last values of $\pi_1, \pi_2$, respectively.
	Let $j$ be the index of the last edge of $T'$ if $T' \ne \emptyset$ and let $j = 0$ otherwise.
	Removing the last values of the two sequences gives a triple with row sets $S_1 \setminus \{u\}$ and $S_2 \setminus \{v\}$ and $|\mathcal Q(T')| = k-c(j, i)$.
	The factor relating the two products is $(-1)^{\mathrm{rank}(S_1 \setminus \{u\}, u) + \mathrm{rank}(S_2 \setminus \{v\}, v)} B[u, e_i] B[v, e_i]$ by \eqref{eq:ses-prefix}.
	Conversely, we reconstruct a triple with last edge $e_i$ by adding $e_i$ to $T'$ and appending $u, v$ to the two shorter sequences.
	The induction hypothesis gives \eqref{eq:ses-dp} by summing over $j$, $u$, and $v$.
\end{proof}

Each term in \eqref{eq:ses-expansion} appears in the sum for exactly one entry $D[i, [n], [n], k]$ and these sums contain no other terms, so
\[
	N_k
	= \sum_{T \in \calT(w, k)}\sum_{\pi_1, \pi_2 \in P_n}v(T, \pi_1, \pi_2)
	= \sum_{i=1}^{m}D[i, [n], [n], k].
\]

Combining the above method for computing $N_k$ with a method for recovering a smallest SES gives the following theorem.

\sesresult*
\begin{proof}
	We evaluate \eqref{eq:ses-dp} in increasing order of $i$ to obtain all $N_k$.
	There are $O(mn4^n)$ states and $O(mn^2)$ terms in each sum in \eqref{eq:ses-dp}.
	Since each $D[i, S_1, S_2, k]$ is a sum of at most $n^{O(n)}$ values in $\{0, \pm1\}$, each value can be represented using $O(n\log(n+1))$ bits and each term and addition in \eqref{eq:ses-dp} can be computed in time polynomial in $n$.
	Since $m = O(n^2)$, all $N_k$ can be computed in $O^*(4^n)$ time and space.
	Taking $k^* = \min\{k \in [0, n] \mid N_k > 0\}$ gives the minimum size $s(w) = k^*$.

	We construct a spanning tree $T$ with $|\mathcal Q(T)| = k^*$ by deciding whether each edge belongs to $T$ in the order $e_1, \dots, e_m$.
	For each $e_\ell$, we compute $N_{k^*}$ under the conditions already added for $e_1, \dots, e_{\ell-1}$ and the additional condition $e_\ell \in T$ using \eqref{eq:ses-dp}.
	We add the condition $e_\ell \in T$ if the resulting value is positive and the condition $e_\ell \notin T$ otherwise.
	Specifically, we apply the two types of conditions as follows.
	For each condition $e_\ell \notin T$, we set every value $D[\ell, S_1, S_2, k]$ to $0$.
	For each condition $e_\ell \in T$, we set to $0$ every term with $j < \ell < i$ in \eqref{eq:ses-dp} and restrict the final sum for $N_{k^*}$ to $i \in [\ell, m]$.
	The latter restrictions ensure that a sequence of chosen edges cannot skip $e_\ell$ or end before it.
	Each choice preserves the existence of a spanning tree $T$ with $|\mathcal Q(T)| = k^*$ satisfying all added conditions.
	After all $m$ choices the edges required to belong to $T$ form such a spanning tree.
	Since we compute $N_{k^*}$ once for each of the $m$ edges, we can compute $T$ in $O^*(4^n)$ time and space.
	Once $T$ is determined, we can construct $\SES{\mathcal Q(T)}$ of size $k^* = s(w)$ in polynomial time.
\end{proof}

\subsection{A Randomized Algorithm}
In the randomized algorithm, we replace $(\det B_T)^2$ with $\det B_T$ so that each state is indexed by a single set $S$ instead of two sets $S_1$ and $S_2$.
However, since $\det B_T$ can be either $1$ or $-1$ for a spanning tree $T$, the sum $\sum_{T \in \calT(w, k)} \det B_T$ may be zero even when a spanning tree with $|\mathcal Q(T)| = k$ exists.
To avoid this problem, we assign an independent random value $y_e$ to each edge $e$ and multiply $\det B_T$ by $\prod_{e \in T} y_e$.

For $k \in [0, n]$, we define the polynomial $R_k$ in the variables $y = (y_e)_{e \in E(w)}$ by
for each edge $e \in E(w)$ and write $y = (y_e)_{e \in E(w)}$.
For $k \in [0, n]$, we define
\begin{equation}
	\label{eq:ses-rand}
	R_k(y) = \sum_{T \in \calT(w, k)} \det(B_T)\prod_{e \in T} y_e.
\end{equation}
Since different edge sets give different monomials and $\det(B_T) \ne 0$ exactly for spanning trees by \cref{lem:tree-det}, the polynomial $R_k$ is nonzero exactly when $N_k > 0$.
For $T = \{e_{j_1}, \dots, e_{j_\ell}\}$ with $j_1 < \dots < j_\ell$ and $0 \le \ell \le n$, let
\[
	v_y(T, \pi) = \sgn(\pi)\prod_{h=1}^{\ell} B[\pi(h), e_{j_h}]y_{e_{j_h}}
\]
for $\pi \in P_{n,\ell}$, with value $1$ for the empty edge set and empty sequence.
By the Leibniz formula, we have $R_k(y) = \sum_{T \in \calT(w, k)} \sum_{\pi \in P_n} v_y(T, \pi)$.

We define $A[i, S, k]$ analogously to $D[i, S_1, S_2, k]$ by summing $v_y(T, \pi)$ over pairs $(T, \pi)$ with $\pi \in P_{n,|T|}$ and $\Set(\pi) = S$.
The conditions on $i$, $T$, and $k$ are unchanged.
We set $A[0, \emptyset, 0] = 1$ and all other values with $i = 0$ or $k \notin [0, n]$ to $0$.
For $i \in [1, m]$, we can compute the value of $A[i, S, k]$ by

\begin{equation}
	\label{eq:ses-rand-dp}
	\begin{aligned}
		A[i, S, k]
		 & = y_{e_i}\sum_{j=0}^{i-1}\sum_{u \in S}
		B[u, e_i](-1)^{\mathrm{rank}(S \setminus \{u\}, u)} \\
		 & \qquad\cdot A[j, S \setminus \{u\}, k-c(j, i)].
	\end{aligned}
\end{equation}
By the same induction as in \cref{lem:ses-dp-correct}, we obtain $R_k(y) = \sum_{i=1}^{m} A[i, [n], k]$.

We assign independent random values to the variables $y_e$ to reduce the probability of cancellation in $R_k(y)$.
The following theorem gives a randomized algorithm that computes a smallest SES with high probability.

\sesrandresult*
\begin{proof}
	Let $M$ be an integer satisfying $2n \le M < 4n$.
	We perform $n$ independent trials by choosing each $y_e$ independently and uniformly from $[M]$ and evaluating all $R_k(y)$ over the integers by \eqref{eq:ses-rand-dp}.
	We take the smallest $k \in [0, n]$ for which $R_k(y) \ne 0$ in at least one trial if such a $k$ exists.
	Since $R_j(y) = 0$ for every $j < k^*$ regardless of the values of $y$, the chosen $k$ satisfies $k \ge k^*$.
	The polynomial $R_{k^*}$ is nonzero of total degree $n$, so the Schwartz--Zippel lemma~\cite[Corollary~1]{Schwartz80} gives $\Pr[R_{k^*}(y) = 0] \le n/M$ in each trial.
	Since the $n$ trials are independent, the chosen $k$ equals $k^*$ with probability at least $1-(n/M)^n \ge 1-2^{-n}$.

	Since $|B[\pi(i), e_{j_i}]y_{e_{j_i}}| \le M$ for each $i \in [|T|]$, we have $|v_y(T, \pi)| \le M^n$.
	Each $A[i, S, k]$ is a sum of at most $n^{O(n)}$ such values, so $|A[i, S, k]| \le n^{O(n)}M^n$.
	By the choice of $M$, each value can be represented using $O(n\log(n+1))$ bits and each term and addition in \eqref{eq:ses-rand-dp} can be computed in time polynomial in $n$.
	There are $O(mn2^n)$ states, and each value can be computed in time polynomial in $n$ using \eqref{eq:ses-rand-dp}.
	Including the time for sampling the random values, all $n$ trials take $O^*(2^n)$ worst-case time and space.

	If all $n$ trials have $R_k(y) = 0$ for every $k \in [0, n]$, we return the SES consisting of the $n$ symbol assignments $(v, w[v])$ for $v \in [n]$.
	Otherwise, let $y$ be the values from a trial with $R_k(y) \ne 0$ for the minimum $k$ obtained above.
	As in the proof of \cref{thm:ses-exact-det}, we determine whether each of $e_1, \dots, e_m$ belongs to $T$ in this order to obtain $T$ with $|\mathcal Q(T)| = k$.
	For each $e_\ell$, we evaluate $R_k(y)$ under the previously fixed conditions and the additional condition $e_\ell \in T$ using \eqref{eq:ses-rand-dp} with the same $y$.
	We restrict the entries $A[i,S,k]$ and the final sum for $R_k(y)$ in the same way as the entries $D[i,S_1,S_2,k]$ and the final sum for $N_k$ in the deterministic algorithm.
	After computing $R_k(y)$ under the restrictions, we add a new condition $e_\ell \in T$ if $R_k(y) \ne 0$ and $e_\ell \notin T$ otherwise.
	We compute $R_k(y)$ at most $m$ times using \eqref{eq:ses-rand-dp} under the fixed conditions, so this procedure takes $O^*(2^n)$ worst-case time and space.
	Once $T$ is determined, we can construct $\SES{\mathcal Q(T)}$ of size $k$ in polynomial time.
\end{proof}
 \section{Approximation Algorithms for Smallest Substring Equation Systems}
\label{sec:approximation_ses}

In this section, we introduce approximation algorithms for the smallest SES problem and show the following theorem.

\sesapproxresult*
The parameter $\alpha$ controls the trade-off between the approximation ratio and the running time.
Setting $\alpha = \ln n$ gives a polynomial-time $(1+\ln n)$-approximation algorithm, while setting $\alpha = 0$ gives exact algorithms with the same time and space bounds as \cref{thm:ses-exact-det} and \cref{thm:ses-exact-rand}.
For the rest of this section, we fix a parameter $\alpha \ge 0$ and let $\tau = \lceil e^{-\alpha}n \rceil$.

Our algorithm consists of two steps.
We first select constraints greedily to reduce the number of connected components in the graph induced by these constraints.
We then contract each connected component and use the exact algorithms from \cref{sec:exact_ses} to find a smallest set of additional constraints representing $w$ together with previously selected constraints.
The running time of the second step depends exponentially on the number of remaining components, so we stop the greedy selection when this number is small enough to meet the desired time bound.

\subsection{Greedy Selection}
\label{sec:ses-greedy}

By \cref{lem:ses-char}, the constraints of a smallest SES representing $w$ can be chosen from the maximal substring equations and symbol assignments corresponding to $\mathcal R(w)$.
For $\mathcal Q \subseteq \mathcal R(w)$, we define $\phi(\mathcal Q) = n+1 - \kappa(E(\mathcal Q))$, where $\kappa(E)$ denotes the number of connected components of $(\{0\} \cup [n], E)$.
$\SES{\mathcal Q}$ represents $w$ if and only if $\phi(\mathcal Q) = n$.

\begin{lemma}
	\label{lem:ses-submodular}
	The function $\phi$ is a monotone nonnegative submodular function.
\end{lemma}
\begin{proof}
	Every graph on $\{0\} \cup [n]$ has at most $n+1$ connected components, so $\phi(\mathcal Q_1) \ge 0$.
	The function is monotone because $\mathcal Q_1 \subseteq \mathcal Q_2$ implies $E(\mathcal Q_1) \subseteq E(\mathcal Q_2)$, and adding edges cannot increase the number of connected components.
	To prove submodularity, we compare the effects of inserting the edges of $R \in \mathcal R(w)$ in the same order into the graphs with edge sets $E(\mathcal Q_1)$ and $E(\mathcal Q_2)$, where $\mathcal Q_1 \subseteq \mathcal Q_2 \subseteq \mathcal R(w)$.
	Whenever an edge merges two components in the graph for $\mathcal Q_2$, it also merges two components in the graph for $\mathcal Q_1$.
	Hence $\phi(\mathcal Q_1 \cup \{R\}) - \phi(\mathcal Q_1) \ge \phi(\mathcal Q_2 \cup \{R\}) - \phi(\mathcal Q_2)$.
	This inequality proves that $\phi$ is submodular.
\end{proof}

For each $R_1 \in \mathcal R(w)$, we execute the following greedy algorithm.
We start with $\mathcal Q_0 = \emptyset$ and add $R_1$ in the first step to obtain $\mathcal Q_1 = \{R_1\}$.
From the second step onward we repeatedly add a run $R_{t+1} \in \mathcal R(w) \setminus \mathcal Q_t$ maximizing $\phi(\mathcal Q_t \cup \{R_{t+1}\})$ to obtain $\mathcal Q_{t+1} = \mathcal Q_t \cup \{R_{t+1}\}$ until $\phi(\mathcal Q_t) \ge n - \tau$.
We define $\mathcal F$ as the family of outputs $\mathcal Q_t$ obtained from the execution for each $R_1 \in \mathcal R(w)$.
The stopping condition ensures $\phi(\mathcal Q) \ge n - \tau$ for every $\mathcal Q \in \mathcal F$.
Each execution terminates after at most $\max\{1, n-\tau\}$ steps because $\phi(\mathcal Q_t)$ increases by at least one in each iteration and $\phi(\mathcal Q_0) = 0$.
Since $|\mathcal R(w)| = O(n^2)$ and each value of $\phi$ can be computed in polynomial time, we have $|\mathcal F| = O(n^2)$ and can compute $\mathcal F$ in polynomial time.

\begin{lemma}
	\label{lem:ses-greedy-selection}
	For every $\mathcal Q^* \subseteq \mathcal R(w)$ representing a smallest SES, there exists $\mathcal Q \in \mathcal F$ satisfying
	$|\mathcal Q \cup \mathcal Q^*| \le (\alpha+1)s(w)$.
\end{lemma}
\begin{proof}
	Consider a set $\mathcal Q^* \subseteq \mathcal R(w)$ representing a smallest SES of size $s = s(w)$.
	For each $t$ with $\phi(\mathcal Q_t) < n$, let $R^*_{t+1} \in \argmax_{R \in \mathcal Q^* \setminus \mathcal Q_t} \phi(\mathcal Q_t \cup \{R\})$.
	Since $\SES{\mathcal Q^*}$ represents $w$, we have $\phi(\mathcal Q_t \cup \mathcal Q^*) = n$.
	By submodularity and $|\mathcal Q^* \setminus \mathcal Q_t| \le s$, we obtain

	\[
		\begin{aligned}
			n - \phi(\mathcal Q_t)
			 & = \phi(\mathcal Q_t \cup \mathcal Q^*) - \phi(\mathcal Q_t)                  \\
			 & \le \sum_{R \in \mathcal Q^* \setminus \mathcal Q_t}
			\bigl(\phi(\mathcal Q_t \cup \{R\}) - \phi(\mathcal Q_t)\bigr)                  \\
			 & \le s\bigl(\phi(\mathcal Q_t \cup \{R^*_{t+1}\}) - \phi(\mathcal Q_t)\bigr),
		\end{aligned}
	\]
	and hence
	$\phi(\mathcal Q_t \cup \{R^*_{t+1}\}) - \phi(\mathcal Q_t) \ge \frac{n - \phi(\mathcal Q_t)}{s}$.

	To obtain a set satisfying the claimed size bound, we consider an element $R^* \in \argmax_{R \in \mathcal Q^*} \phi(\{R\})$ and analyze the above greedy algorithm with $R_1 = R^*$.
	At $t = 0$ both $R_1 = R^*$ and $R^*_1$ maximize $\phi(\{R\})$ over $R \in \mathcal Q^*$.
	For $t \ge 1$, the run $R_{t+1}$ maximizes $\phi(\mathcal Q_t \cup \{R\})$ over $\mathcal R(w) \setminus \mathcal Q_t$.
	On the other hand, the run $R^*_{t+1}$ maximizes the same quantity over $\mathcal Q^* \setminus \mathcal Q_t$, which is a subset of $\mathcal R(w) \setminus \mathcal Q_t$.
	Thus, every step satisfies

	\[
		\begin{aligned}
			\phi(\mathcal Q_{t+1}) - \phi(\mathcal Q_t)
			 & = \phi(\mathcal Q_t \cup \{R_{t+1}\}) - \phi(\mathcal Q_t)     \\
			 & \ge \phi(\mathcal Q_t \cup \{R^*_{t+1}\}) - \phi(\mathcal Q_t) \\
			 & \ge \frac{n - \phi(\mathcal Q_t)}{s}.
		\end{aligned}
	\]
	Rearranging the above inequality gives
	\[
		\begin{aligned}
			n - \phi(\mathcal Q_{t+1})
			 & = n - \phi(\mathcal Q_t) - \bigl(\phi(\mathcal Q_{t+1}) - \phi(\mathcal Q_t)\bigr) \\
			 & \le \left(1-\frac1s\right)\bigl(n - \phi(\mathcal Q_t)\bigr)                       \\
			 & \le e^{-1/s}\bigl(n - \phi(\mathcal Q_t)\bigr).
		\end{aligned}
	\]
	Since $\phi(\mathcal Q_0) = 0$, repeated application yields $n - \phi(\mathcal Q_t) \le ne^{-t/s}$.

	The algorithm stops when $n - \phi(\mathcal Q_t) \le \tau$.
	Since $n - \phi(\mathcal Q_t) \le ne^{-t/s}$ and $ne^{-\alpha} \le \tau$, the stopping condition holds whenever $t \ge \alpha s$.
	Therefore, we obtain the resulting set $\mathcal Q$ after at most $t_{\max} = \max\{1, \lceil \alpha s\rceil\}$ iterations.
	Since $R^*$ belongs to both $\mathcal Q$ and $\mathcal Q^*$, we have $|\mathcal Q \cup \mathcal Q^*| = |\mathcal{Q}^*| + |\mathcal Q \setminus \mathcal Q^*| \le s+t_{\max}-1 = s + \max\{0, \lceil \alpha s\rceil - 1\} \le (\alpha+1)s$.
\end{proof}

\subsection{Faster Exact Algorithms under Additional Constraints}
\label{sec:ses-additional}

In this section, we fix a set $\mathcal A \subseteq \mathcal R(w)$ and show how to compute a smallest SES representing $w$ that contains the constraints corresponding to all runs in $\mathcal A$.
We denote its size by $s(w \mid \mathcal A)$ and give algorithms whose time and space complexity depend on the number $d = \kappa(E(\mathcal A))$ of connected components.

We number the connected components of $(\{0\} \cup [n], E(\mathcal A))$ by $0, 1, \dots, d-1$ with $0$ assigned to the component containing vertex $0$.
For each vertex $v \in \{0\} \cup [n]$, let $\operatorname{idx}_{\mathcal A}(v)$ be the number of the component containing $v$.
We define $V_{\mathcal A} = \{0\} \cup [d-1]$ and let $e'_i$ be an edge with endpoints $\operatorname{idx}_{\mathcal A}(p)$ and $\operatorname{idx}_{\mathcal A}(q)$ for each original edge $e_i = \{p, q\}$.
We define the undirected multigraph $G_{\mathcal A} = (V_{\mathcal A}, E_{\mathcal A})$ obtained from $G(w)$ by contracting each component, where $E_{\mathcal A} = \{e'_i \mid i \in [|E(w)|]\}$.
We use the same order $e'_1, \dots, e'_{|E(w)|}$ and the same values $c(j, i)$ as in \cref{sec:exact_ses}.
For $T \subseteq E_{\mathcal A}$, we define $\mathcal Q(T) \subseteq \mathcal R(w)$ as the set of runs containing an original edge $e_i$ with $e'_i \in T$.

For $\mathcal A \subseteq \mathcal Q \subseteq \mathcal R(w)$, $\SES{\mathcal Q}$ represents $w$ exactly when the edges $\{e'_i \mid e_i \in E(\mathcal Q \setminus \mathcal A)\}$ connect $G_{\mathcal A}$.
Such an edge set contains a spanning tree $T$ with $\mathcal Q(T) \subseteq \mathcal Q \setminus \mathcal A$.
Conversely, for any spanning tree $T$ of $G_{\mathcal A}$, $\SES{\mathcal A \cup \mathcal Q(T)}$ represents $w$.
For each edge $e_i \in E(\mathcal A)$, its endpoints belong to the same connected component of $(\{0\} \cup [n], E(\mathcal A))$ and the corresponding edge $e'_i$ has both endpoints at the same vertex of $G_{\mathcal A}$.
Thus, no spanning tree $T$ of $G_{\mathcal A}$ contains such an edge, so $\mathcal A \cap \mathcal Q(T) = \emptyset$.
Therefore,
\begin{equation}
	\label{eq:ses-additional-cost}
	s(w \mid \mathcal A) = |\mathcal A| + \min\{|\mathcal Q(T)| \mid T\text{ is a spanning tree of }G_{\mathcal A}\}.
\end{equation}

We define the reduced incidence matrix $B_{\mathcal A} \in \mathbb Z^{[d-1] \times E(w)}$ of $G_{\mathcal A}$ with the column indexed by $e_i$ corresponding to $e'_i$.
If $e'_i$ has both endpoints at the same vertex, then the corresponding column of $B_{\mathcal A}$ is the zero vector.
We give the precise definition of $B_{\mathcal A}$ in Appendix~\ref{sec:ses-fixed-dp}.
For each $T \subseteq E_{\mathcal A}$ with $|T| = d-1$, let $B_{\mathcal A,T}$ consist of the columns corresponding to $T$.
The same argument as in \cref{lem:tree-det} gives $\det B_{\mathcal A,T} = \pm1$ if $T$ is a spanning tree of $G_{\mathcal A}$ and $\det B_{\mathcal A,T} = 0$ otherwise.

For $k \in [0, d-1]$, let
\begin{align*}
	N_k    & = \sum_{\substack{T \subseteq E_{\mathcal A} \\ |T| = d-1,\ |\mathcal Q(T)| = k}} (\det B_{\mathcal A,T})^2,                        \\
	R_k(y) & = \sum_{\substack{T \subseteq E_{\mathcal A} \\ |T| = d-1,\ |\mathcal Q(T)| = k}} \det B_{\mathcal A,T} \prod_{e'_i \in T} y_{e_i}.
\end{align*}
The value $N_k$ counts the spanning trees of $G_{\mathcal A}$ using $k$ runs.
As in \cref{sec:exact_ses}, distinct edge sets give distinct monomials in $R_k$ and hence $R_k$ is nonzero exactly when $N_k > 0$.

We also define the tables $D[i, S_1, S_2, k]$ and $A[i, S, k]$ as in \cref{sec:exact_ses} with $B_{\mathcal A}$ in place of $B$.
The edge set $T$ is now a subset of $E_{\mathcal A}$ and each set of row indices is a subset of $[d-1]$.
The index $i$ specifies the last chosen edge $e'_i$ and $k$ counts the runs containing chosen edges.
Appendix~\ref{sec:ses-fixed-dp} gives the precise definitions and recurrences together with a proof of the following lemma.

\begin{restatable}{lemma}{sesfixeddpcorrect}
	\label{lem:ses-fixed-dp-correct}
	For $d \ge 2$, the initial values and recurrences correctly compute $D[i, S_1, S_2, k]$ and $A[i, S, k]$.
	For every $k \in [0, d-1]$, we have $N_k = \sum_{i=1}^{|E(w)|} D[i, [d-1], [d-1], k]$ and $R_k(y) = \sum_{i=1}^{|E(w)|} A[i, [d-1], k]$.
\end{restatable}
\begin{proof}
	See Appendix~\ref{sec:ses-fixed-dp}.
\end{proof}

The following lemma shows that we can compute $s(w \mid \mathcal A)$ and a corresponding SES using the contracted graph $G_{\mathcal A}$.

\begin{lemma}
	\label{lem:ses-additional}
	Given $w$ and $\mathcal A \subseteq \mathcal R(w)$ with $d = \kappa(E(\mathcal A))$, the value $s(w \mid \mathcal A)$ and an SES of this size representing $w$ that contains the constraints corresponding to $\mathcal A$ can be computed deterministically in $O^*(4^{d})$ time and space, or in $O^*(2^{d})$ worst-case time and space with probability at least $1-2^{-n}$.
\end{lemma}
\begin{proof}
	If $d = 1$, we return $\SES{\mathcal A}$.
	For $d \ge 2$, we compute all $N_k$ using \eqref{eq:ses-fixed-dp} in Appendix~\ref{sec:ses-fixed-dp} and \cref{lem:ses-fixed-dp-correct}.
	Each value $D[i, S_1, S_2, k]$ is a sum of at most $n^{O(d)}$ terms in $\{0, \pm1\}$, so each value can be represented using $O(d\log(n+1))$ bits.
	Each arithmetic operation therefore takes time polynomial in $n$.
	The index $i$ has $|E(w)|+1$ choices and $k$ has $d$ choices.
	Since there are $4^{d-1}$ pairs of subsets $S_1, S_2 \subseteq [d-1]$, the table has at most $(|E(w)|+1)d4^{d-1}$ entries.
	Each recurrence has $O(|E(w)|d^2)$ terms.
	Since $|E(w)| = O(n^2)$ and $d \le n+1$, all $N_k$ can be computed in $O^*(4^{d})$ time and space.
	By \eqref{eq:ses-additional-cost}, the smallest $k$ with $N_k > 0$ equals $s(w \mid \mathcal A) - |\mathcal A|$.

	For the randomized algorithm, we choose an integer $M$ satisfying $2n \le M < 4n$ as in the proof of \cref{thm:ses-exact-rand}.
	We perform $n$ independent trials with each $y_e$ chosen independently and uniformly from $[M]$.
	By the same argument as in that proof, the smallest $k$ satisfying $R_k(y) \ne 0$ in at least one of the $n$ trials equals $s(w \mid \mathcal A) - |\mathcal A|$ with probability at least $1-((d-1)/M)^n \ge 1-2^{-n}$.
	Each value $A[i, S, k]$ is a sum of at most $n^{O(d)}$ terms of absolute value at most $M^{d-1}$, so it can be represented using $O(d\log(n+1))$ bits and each arithmetic operation takes time polynomial in $n$.
	There are at most $(|E(w)|+1)d2^{d-1}$ entries and each recurrence has $O(|E(w)|d)$ terms.
	These bounds show that the $n$ trials can be performed in $O^*(2^{d})$ worst-case time and space.

	We recover a spanning tree $T$ for the computed minimum $k$ within these time and space bounds using the methods in the proofs of \cref{thm:ses-exact-det} and \cref{thm:ses-exact-rand}.
	The SES $\SES{\mathcal A \cup \mathcal Q(T)}$ has size $|\mathcal A| + k$ and represents $w$ with all constraints of $\SES{\mathcal A}$.
\end{proof}

\subsection{Putting Everything Together}

By combining the greedy algorithm from \cref{sec:ses-greedy} with the exact algorithms from \cref{sec:ses-additional}, we prove the main theorem of this section.

\begin{proof}[Proof of \cref{thm:ses-approximation}]
	We first construct a family $\mathcal F$ using the greedy algorithm from \cref{sec:ses-greedy} with $|\mathcal F| = O(n^2)$ and $\phi(\mathcal A) \ge n - \tau$ for every $\mathcal A \in \mathcal F$.
	For each $\mathcal A \in \mathcal F$, we compute an SES containing the constraints corresponding to $\mathcal A$ by \cref{lem:ses-additional} and return a smallest SES among these outputs.

	Let $\mathcal Q^* \subseteq \mathcal R(w)$ be a set of runs representing a smallest SES.
	By \cref{lem:ses-greedy-selection}, there exists $\mathcal Q \in \mathcal F$ such that $|\mathcal Q \cup \mathcal Q^*| \le (\alpha+1)s(w)$.
	Since $\SES{\mathcal Q \cup \mathcal Q^*}$ represents $w$ and contains all constraints of $\SES{\mathcal Q}$, we have $s(w \mid \mathcal Q) \le |\mathcal Q \cup \mathcal Q^*|$.
	Since the deterministic algorithm returns an SES of size $s(w \mid \mathcal A)$ for some $\mathcal A \in \mathcal F$ minimizing this value, we obtain
	\[
		s(w \mid \mathcal A)
		= \min_{\mathcal A' \in \mathcal F} s(w \mid \mathcal A') \le s(w \mid \mathcal Q) \le |\mathcal Q \cup \mathcal Q^*| \le (\alpha+1)s(w).
	\]
	For the randomized algorithm, it suffices that the SES computed for $\mathcal Q$ has the minimum size $s(w \mid \mathcal Q)$.
	In this case, the returned SES has size at most $s(w \mid \mathcal Q)$ and satisfies the same approximation bound.
	By \cref{lem:ses-additional}, this holds with probability at least $1-2^{-n}$.

	The greedy algorithm from \cref{sec:ses-greedy} constructs $\mathcal F$ in polynomial time.
	Its stopping condition ensures $\kappa(E(\mathcal A)) \le \tau+1 \le e^{-\alpha}n + 2$ for every $\mathcal A \in \mathcal F$,
	so the parameter $d = \kappa(E(\mathcal A))$ used in \cref{sec:ses-additional} is at most $e^{-\alpha}n + 2$.
	By \cref{lem:ses-additional}, each SES can be computed in $O^*(4^{e^{-\alpha}n + 2}) = O^*(4^{e^{-\alpha}n})$ deterministic or $O^*(2^{e^{-\alpha}n + 2}) = O^*(2^{e^{-\alpha}n})$ randomized worst-case time and space.
	Since $|\mathcal F| = O(n^2)$, computing an SES for every $\mathcal A \in \mathcal F$ preserves these bounds.
\end{proof}

It is known that $s(w) \le b(w)$ and that an SES of size $k$ representing $w$ can be converted in polynomial time into a BMS of size at most $4k$~\cite{ShibataB26}.
Combining these results with \cref{thm:ses-approximation} gives the following approximation for the smallest BMS.

\begin{corollary}
	\label{cor:bms-approximation}
	For every $\alpha \ge 0$, a BMS for $w$ of size at most $4(\alpha + 1)b(w)$ can be computed deterministically in $O^*(4^{e^{-\alpha} n})$ time and space, or with probability at least $1 - 2^{-n}$ in $O^*(2^{e^{-\alpha} n})$ worst-case time and space.
\end{corollary}

\section{Smallest Bidirectional Macro Schemes}
\label{sec:exact_bms}

In this section, we give a deterministic exact algorithm for computing a smallest BMS.

\bmsresult*

\subsection{Characterization of the Minimum BMS}

As in~\Cref{sse:char-minSES}, we reduce the problem of finding a smallest BMS to an optimization problem over vectors representing valid reference functions.

We represent a reference function $\varphi$ by the vector $\varphi = (\varphi(1), \dots, \varphi(n))^{\mathsf T} \in (\{0\} \cup [n])^n$.
For $i\in[n]$, let $E_i=\{0\}\cup\{j\in[n]\setminus\{i\}\mid w[j]=w[i]\}$.
A vector $\varphi \in (\{0\} \cup [n])^n$ is \emph{symbol-respecting} if $\varphi(i)\in E_i$ for every $i\in[n]$, and let $\Phi(w)$ be the set of symbol-respecting vectors.
We use the same definition of validity for reference functions and their vector representations.
By definition, every valid vector $\varphi$ is symbol-respecting.
We define the set of valid vectors as $\Phi_{\mathsf{valid}}(w) \subseteq \Phi(w)$.
For $x, y \in \{0\} \cup [n]$, let $c(x, y) = 1$ if $x \ne 0$ and $y = x + 1$, and $0$ otherwise.
For a sequence $A = (x_1, \dots, x_m)$ of such values, we define $C(A) = \sum_{i=1}^{m-1} c(x_i, x_{i+1})$, with $C(A) = 0$ for the empty sequence.

If $c(\varphi(i), \varphi(i+1)) = 1$, the adjacent positions $i$ and $i+1$ refer to consecutive source positions and can be copied together.
Thus $C(\varphi)$ counts the phrase boundaries that can be omitted.
The following lemma expresses $b(w)$ by maximizing the number of omitted boundaries over vectors in $\Phi_{\mathsf{valid}}(w)$.

\begin{lemma}
	\label{lem:bms-char}
	$b(w) = n - \max\{C(\varphi) \mid \varphi \in \Phi_{\mathsf{valid}}(w)\}$.
\end{lemma}
\begin{proof}
	($\ge$) Consider a valid BMS with $m$ phrases of lengths $\ell_1, \dots, \ell_m$.
	Its reference function corresponds to a valid vector $\varphi \in \Phi_{\mathsf{valid}}(w)$.
	If positions $i$ and $i+1$ belong to the same phrase, then $c(\varphi(i), \varphi(i+1)) = 1$.
	Hence $C(\varphi) \ge \sum_{q=1}^m (\ell_q - 1) = n - m$.

	($\le$) Let $\varphi\in\Phi_{\mathsf{valid}}(w)$.
	We split $[n]$ into $n - C(\varphi)$ blocks of consecutive positions such that $i$ and $i+1$ lie in the same block exactly when $c(\varphi(i), \varphi(i+1)) = 1$.
	A block $[l,r]$ with $r>l$ satisfies $\varphi(l)>0$ and $\varphi(l+h)=\varphi(l)+h$ for $0\le h\le r-l$, so $w[l..r]=w[\varphi(l)..\varphi(l)+r-l]$ and $\varphi(l)\ne l$, and we make it a copy phrase with source $\varphi(l)$.
	For each singleton block, we store its symbol explicitly.
	Changing the references at singleton blocks to $0$ cannot create a directed cycle.
	This shows that the resulting BMS is valid, and it has $n - C(\varphi)$ phrases.
\end{proof}

\subsection{Checking Validity by Matrix Determinants}

As in \cref{sec:ses-deterministic}, we characterize the validity of a BMS by the determinant of the corresponding matrix.

For $\varphi \in \Phi(w)$, let $G_{\varphi}$ be the directed graph with vertex set $\{0\} \cup [n]$ and arcs $i \to \varphi(i)$ for all $i \in [n]$.
The graph $G_{\varphi}$ is a directed tree with all arcs directed toward the root $0$ if $\varphi$ is valid and contains a directed cycle otherwise.

We define $M_{\varphi}\in\mathbb Z^{n\times n}$ by

\begin{equation}
	\label{eq:bms-matrix}
	M_{\varphi}[i,j]=\begin{cases}
		1  & \text{if }i=j,          \\
		-1 & \text{if }j=\varphi(i), \\
		0  & \text{otherwise}.
	\end{cases}
\end{equation}

By definition, $M_{\varphi}$ is obtained by deleting the column corresponding to $0$ from the incidence matrix of $G_{\varphi}$ with rows indexed by arcs $i \to \varphi(i)$ and columns indexed by vertices.

Unlike in the SES case, $M_{\varphi}$ directly corresponds to the incidence matrix of the directed graph $G_{\varphi}$.
This property allows us to characterize validity by the determinant itself rather than its absolute value, as stated in the following lemma.
\begin{lemma}
	\label{lem:det}
	If $\varphi$ is valid, then $\det M_{\varphi}=1$. Otherwise, $\det M_{\varphi}=0$.
\end{lemma}
\begin{proof}
	If $\varphi$ is invalid, the graph $G_{\varphi}$ contains a directed cycle $i_1\to i_2\to\dots\to i_q\to i_1$.
	The sum of the rows of $M_{\varphi}$ corresponding to $i_1,\dots,i_q$ is the zero vector, so these rows are linearly dependent and hence $\det M_{\varphi}=0$.

	If $\varphi$ is valid, then $G_{\varphi}$ is a directed tree rooted at $0$, so there is an ordering of its vertices with $0$ last in which every arc points from an earlier vertex to a later vertex.
	We reorder the rows and columns of $M_{\varphi}$ according to this vertex ordering restricted to $[n]$.
	Let $M'_{\varphi}$ denote the resulting matrix.
	Since every arc points to a later vertex in this ordering, $M'_{\varphi}$ is an upper triangular matrix.
	All diagonal entries of $M'_{\varphi}$ are $1$ by definition, and the determinant of a triangular matrix is the product of its diagonal entries, so $\det M'_{\varphi}=1$.
	Permuting the rows and columns in the same way does not change the determinant, thus we have $\det M_{\varphi}=\det M'_{\varphi}=1$.
\end{proof}

\begin{example}
	Let $w=\mathtt{abaa}$ and $\varphi_1=(3,0,0,1)^{\mathsf T}$.
	The graph $G_{\varphi_1}$ has arcs $1 \to 3$, $2 \to 0$, $3 \to 0$, and $4 \to 1$, so it is a directed tree rooted at $0$.
	We reorder the rows and columns of $M_{\varphi_1}$ according to the vertex ordering $(4,1,3,2)$ and denote the resulting matrix by $M'_{\varphi_1}$:

	\[
		M_{\varphi_1}=\begin{pmatrix}
			1  & 0 & -1 & 0 \\
			0  & 1 & 0  & 0 \\
			0  & 0 & 1  & 0 \\
			-1 & 0 & 0  & 1
		\end{pmatrix},\qquad
		M'_{\varphi_1}=\begin{pmatrix}
			1 & -1 & 0  & 0 \\
			0 & 1  & -1 & 0 \\
			0 & 0  & 1  & 0 \\
			0 & 0  & 0  & 1
		\end{pmatrix}.
	\]

	The matrix $M'_{\varphi_1}$ is upper triangular and all diagonal entries are $1$, so $\det M_{\varphi_1}=\det M'_{\varphi_1}=1$.

	For the same string $w$, let $\varphi_2=(3,0,4,1)^{\mathsf T}$.
	The graph $G_{\varphi_2}$ contains the directed cycle $1\to3\to4\to1$, and

	\[
		M_{\varphi_2}=\begin{pmatrix}
			1  & 0 & -1 & 0  \\
			0  & 1 & 0  & 0  \\
			0  & 0 & 1  & -1 \\
			-1 & 0 & 0  & 1
		\end{pmatrix}.
	\]

	The sum of rows $1$, $3$, and $4$ of $M_{\varphi_2}$ is the zero vector, so $\det M_{\varphi_2}=0$.
\end{example}

\subsection{Exact Algorithm for Smallest BMSs}

As in \eqref{eq:ses-expansion} in \cref{sec:ses-deterministic}, we define $N_k = \sum_{\varphi \in \Phi(w, k)} \det M_{\varphi}$ for $k \in [0, n-1]$ using $\Phi(w, k) = \{\varphi \in \Phi(w) \mid C(\varphi) = k\}$.
By \cref{lem:det}, $N_k$ counts the valid vectors in $\Phi(w, k)$, so it suffices to find the largest $k$ with $N_k > 0$.

For $0 \le t \le n$ and an integer $k$, let $\Phi^{(t)}(w, k)$ be the set of vectors $\varphi = (\varphi(1), \dots, \varphi(t))^{\mathsf T}$ such that $\varphi(i) \in E_i$ for every $i \in [t]$ and $C(\varphi) = k$.
In particular, $\Phi^{(n)}(w, k) = \Phi(w, k)$.
For $t = 0$, the empty vector is the unique member of $\Phi^{(0)}(w, 0)$, and $\Phi^{(0)}(w, k) = \emptyset$ for $k \ne 0$.

We extend the definition of $M_{\varphi}$ to vectors $\varphi \in \Phi^{(t)}(w, k)$ by defining the $t \times n$ matrix $M_{\varphi}$ according to \eqref{eq:bms-matrix} for $i \in [t]$ and $j \in [n]$.
For $\pi \in P_{n,t}$, we define

\[
	v(\varphi, \pi) = \sgn(\pi) \prod_{i=1}^{t} M_{\varphi}[i, \pi(i)].
\]

Since $\pi \in P_{n,t}$, the factors $M_{\varphi}[i, \pi(i)]$ for $i \in [t]$ belong to distinct columns.
By the definition of $M_{\varphi}$, each row $i$ has nonzero values only in columns $\{i, \varphi(i)\} \setminus \{0\}$.
If this condition fails for some row, then $v(\varphi, \pi) = 0$.
By the Leibniz formula for the determinant, we have

\begin{equation}
	\label{eq:bms-expansion}
	N_k
	= \sum_{\varphi \in \Phi(w, k)} \det M_{\varphi}
	= \sum_{\varphi \in \Phi(w, k)} \sum_{\pi \in P_n} v(\varphi, \pi).
\end{equation}

To compute each term $v(\varphi, \pi)$ efficiently, we compute the products for the first $t$ entries of the reference vector and column sequence in increasing order of $t$.
For $t \ge 1$, let $\varphi'$ and $\pi'$ be obtained by deleting the last entries of $\varphi \in \Phi^{(t)}(w, k)$ and $\pi \in P_{n,t}$, respectively.

The sequences $\pi$ and $\pi'$ satisfy $\mathrm{inv}(\pi) - \mathrm{inv}(\pi') = \mathrm{rank}(\Set(\pi'), \pi(t))$ because they have the same first $t-1$ entries.
Since $M_{\varphi}$ and $M_{\varphi'}$ agree on their first $t-1$ rows, we have $\prod_{i=1}^{t} M_{\varphi}[i, \pi(i)] = \left(\prod_{i=1}^{t-1} M_{\varphi'}[i, \pi'(i)]\right) \cdot M_{\varphi}[t, \pi(t)]$.
Combining the changes in the sign and the product gives

\begin{equation}
	\label{eq:bms-prefix}
	v(\varphi, \pi)
	= v(\varphi', \pi') \cdot (-1)^{\mathrm{rank}(\Set(\pi'), \pi(t))} \cdot M_{\varphi}[t, \pi(t)].
\end{equation}

The value of $M_{\varphi}[t, \pi(t)]$ is determined by $\varphi(t)$ and $\pi(t)$.
Thus, the factor multiplying $v(\varphi', \pi')$ in \eqref{eq:bms-prefix} depends only on $\Set(\pi')$, $\varphi(t)$, and $\pi(t)$.
The sequences $\varphi$ and $\pi$ must belong to $\Phi^{(t)}(w, k)$ and $P_{n,t}$, respectively.
For $\varphi$, each new entry must satisfy $\varphi(t) \in E_t$.
For $\pi$, each new entry must satisfy $\pi(t) \in [n] \setminus \Set(\pi')$.
To compute $N_k$ for each $k$, we also keep the value $C(\varphi)$ for each prefix.
For $t \ge 2$, the value $C(\varphi)$ can be computed by $C(\varphi) = C(\varphi') + c(\varphi'(t-1), \varphi(t))$, so we store the last value $\varphi(t)$ and the current value $C(\varphi)$ to update $C$.
For $S \subseteq [n]$ with $t = |S| \ge 1$, we consider sequences $\varphi \in \Phi^{(t)}(w, k)$ and $\pi \in P_{n,t}$ and define

\begin{align*}
	\mathcal{D}(S, x, k)
	 & = \{(\varphi, \pi) \in \Phi^{(t)}(w, k) \times P_{n,t}
	\mid \Set(\pi) = S,\ \varphi(t) = x\},                               \\
	D[S, x, k]
	 & = \sum_{(\varphi, \pi) \in \mathcal{D}(S, x, k)} v(\varphi, \pi).
\end{align*}

We use the initial values $D[\emptyset, 0, 0] = 1$ and $D[\emptyset, x, k] = 0$ for $(x, k) \ne (0, 0)$.
For nonempty $S$ with $t = |S|$, we set $D[S, x, k] = 0$ if $x \not\in E_t$ or $k < 0$.
For $x \in E_t$ and $k \ge 0$, we compute $D[S, x, k]$ by the recurrence

\begin{equation}
	\label{eq:bms-dp}
	D[S, x, k]
	= \sum_{y=0}^{n}
	\sum_{\substack{(j, e) \in \{(t, 1), (x, -1)\} \\ j \in S}}
	D[S \setminus \{j\}, y, k - c(y, x)]
	\cdot e \cdot (-1)^{\mathrm{rank}(S \setminus \{j\}, j)}.
\end{equation}
Here, $y$ corresponds to the last value of the prefix of length $t-1$, with $y = 0$ for the empty prefix.
The choice $(j, e) = (t, 1)$ corresponds to the diagonal entry $M_{\varphi}[t, t] = 1$ of row $t$, and $(j, e) = (x, -1)$ corresponds to the entry $M_{\varphi}[t, x] = -1$.
If $x = 0$, the latter choice is excluded by $j \in S$.
Each term refers to prefixes of length $t-1$ with the value $C(\varphi') = k - c(y, x)$.

The following lemma shows the correctness of \eqref{eq:bms-dp}.
\begin{restatable}{lemma}{bmsdpinvariant}
	\label{lem:dp-inv}
	The initial values given above and the recurrence in \eqref{eq:bms-dp} correctly compute $D[S, x, k]$ for every nonempty $S \subseteq [n]$.
\end{restatable}
\begin{proof}
	We prove the claim by induction on $t$.
	For $t = 0$, there is one pair of empty sequences with $v(\varphi, \pi) = 1$ and $C(\varphi) = 0$, as specified by the initial values with $x = 0$.

	For the induction step, let $S$ be nonempty and $t = |S|$.
	If $x \not\in E_t$ or $k < 0$, then $\mathcal{D}(S, x, k)$ is empty and $D[S, x, k] = 0$.
	Otherwise, consider a pair $(\varphi, \pi) \in \mathcal{D}(S, x, k)$ and let $j = \pi(t)$.
	If $M_{\varphi}[t, j] = 0$, its value $v(\varphi, \pi)$ is $0$.
	If $M_{\varphi}[t, j] \ne 0$, then $(j, e) \in \{(t, 1), (x, -1)\}$ with $e = M_{\varphi}[t, j]$ and $j \in S$.
	Let $y = \varphi(t-1)$ when $t \ge 2$, and $y = 0$ when $t = 1$.
	Removing the last value of each sequence leaves a pair $(\varphi', \pi')$ with column set $S \setminus \{j\}$ and last value $y$.
	The shorter prefix satisfies $C(\varphi') = k - c(y, x)$, where $c(0, x) = 0$ also covers $t = 1$.
	By \eqref{eq:bms-prefix} and the induction hypothesis, summing over the pairs in $\mathcal{D}(S, x, k)$ gives \eqref{eq:bms-dp}.
\end{proof}

For $S = [n]$, every pair $(\varphi, \pi) \in \Phi(w, k) \times P_n$ belongs to exactly one $\mathcal D([n], x, k)$ with $x \in E_n$, so for every $k \in [0, n-1]$ we have
$N_k
	= \sum_{\varphi \in \Phi(w, k)}\sum_{\pi \in P_n}v(\varphi, \pi)
	= \sum_{x \in E_n}D[[n], x, k]$.
Thus, we can compute all $N_k$ and $b(w)$ using the values $D[S, x, k]$ obtained from \eqref{eq:bms-dp}.

Combining the computation of $b(w)$ with a method for reconstructing a valid vector $\varphi$ maximizing $C(\varphi)$ gives the following theorem.
\bmsresult*

\begin{proof}
	We compute $N_k$ for each $k$ using the recurrence for $D[S, x, k]$ in \eqref{eq:bms-dp} and the above identity, and obtain $b(w) = n - k^*$, where $k^* = \max\{k \in [0, n-1] \mid N_k > 0\}$.
	The total number of states is $O(n^2 2^n)$.
	Each sum in \eqref{eq:bms-dp} has at most $2n$ nonzero terms, and the indices and sign of each term can be computed in $O(n)$ time.
	We next bound the absolute value of $D[S, x, k]$ to determine the number of bits needed for its representation.
	For $t = |S|$, each of $\varphi$ and $\pi$ has at most $n^t$ choices, so $|\mathcal{D}(S, x, k)| \le n^{2t} \le n^{2n}$.
	Since $|v(\varphi, \pi)| \le 1$, we have $|D[S, x, k]| \le n^{2n}$, and each value can be represented using $O(n \log(n+1))$ bits.
	Thus, each term and addition in \eqref{eq:bms-dp} can be computed in time polynomial in $n$.
	By computing the values in increasing order of $|S|$, we obtain all $N_k$ in $O^*(2^n)$ time and space.

	To construct a smallest BMS, we determine the values $\varphi(1), \dots, \varphi(n)$ in this order so that $\varphi$ is valid and $C(\varphi) = k^*$.
	Suppose that the first $t-1$ values of such a vector have been determined.
	For each candidate $y \in E_t$ for $\varphi(t)$, we compute the values $D[S, x, k]$ by \eqref{eq:bms-dp} with $\varphi(1), \dots, \varphi(t-1)$ fixed to these values and $\varphi(t)$ fixed to $y$.
	Specifically, we set every value $D[S, x, k]$ with $1 \le |S| \le t$ and $x \ne \varphi(|S|)$ to $0$.
	This gives $N'_k$, the number of valid vectors with the fixed values of $\varphi(1), \dots, \varphi(t)$ and $C(\varphi) = k$.
	If some $N'_k$ is positive, the largest $k$ with $N'_k > 0$ is the maximum value of $C(\varphi)$ under these restrictions.
	We choose $y$ as $\varphi(t)$ if this maximum equals $k^*$, the largest $k$ with $N_k > 0$ when no values of $\varphi$ are fixed.
	Such a candidate exists at every step because each previous choice preserves a valid vector with $C(\varphi) = k^*$.
	After determining all $n$ values, we have a valid vector $\varphi$ with $C(\varphi) = k^*$.
	We compute all $N'_k$ for at most $n^2$ pairs $(t, y)$, so determining $\varphi$ takes $O^*(2^n)$ time and space.
	From $\varphi$, we construct a BMS with $n - C(\varphi) = b(w)$ phrases in $O(n)$ time as in the proof of \cref{lem:bms-char}.
\end{proof}

\section*{AI Usage Disclosure}
\label{sec:ai_usage_disclosure}
The authors used OpenAI's ChatGPT and Anthropic's Claude to obtain the main results\\(\cref{thm:ses-exact-det,thm:ses-exact-rand,thm:ses-approximation,thm:bms}) and their proofs.
These models were also used for drafting and correcting the text, improving its presentation, and preparing and editing the figure files.
The authors verified all claims and proofs and take full responsibility for the contents of this paper.
 
\clearpage
\bibliography{refs}

@article{CharikarLLPPSS05,
  author    = {Moses Charikar and Eric Lehman and Ding Liu and Rina Panigrahy and Manoj Prabhakaran and Amit Sahai and abhi shelat},
  title     = {The Smallest Grammar Problem},
  journal   = {IEEE Transactions on Information Theory},
  volume    = {51},
  number    = {7},
  pages     = {2554--2576},
  year      = {2005},
  doi       = {10.1109/TIT.2005.850116}
}

@article{CaselFGGS21,
  author    = {Katrin Casel and Henning Fernau and Serge Gaspers and Benjamin Gras and Markus L. Schmid},
  title     = {On the Complexity of the Smallest Grammar Problem over Fixed Alphabets},
  journal   = {Theory of Computing Systems},
  volume    = {65},
  number    = {2},
  pages     = {344--409},
  year      = {2021},
  doi       = {10.1007/s00224-020-10013-w}
}

@inproceedings{KawamotoIKB24,
  author    = {Akiyoshi Kawamoto and Tomohiro I and Dominik K{\"o}ppl and Hideo Bannai},
  title     = {On the Hardness of Smallest {RLSLPs} and Collage Systems},
  booktitle = {2024 Data Compression Conference (DCC)},
  pages     = {243--252},
  publisher = {IEEE},
  year      = {2024},
  doi       = {10.1109/DCC58796.2024.00032}
}

@inproceedings{BanbaraBHKMN26,
  author    = {Mutsunori Banbara and Hideo Bannai and Takashi Horiyama and Dominik K{\"o}ppl and Takuya Mieno and Hidetomo Nabeshima},
  title     = {Optimal Dictionary-Based Compression with Answer Set Programming: Encodings and Empirical Analysis},
  booktitle = {23rd International Conference on Principles of Knowledge Representation and Reasoning (KR 2026)},
  pages     = {700--710},
  publisher = {International Joint Conferences on Artificial Intelligence Organization},
  year      = {2026},
  doi       = {10.24963/kr.2026/66}
}

@article{StorerS82,
  author    = {James A. Storer and Thomas G. Szymanski},
  title     = {Data compression via textual substitution},
  journal   = {Journal of the ACM},
  volume    = {29},
  number    = {4},
  pages     = {928--951},
  year      = {1982},
  doi       = {10.1145/322344.322346}
}

@inproceedings{BannaiGIKKN22,
  author    = {Hideo Bannai and Keisuke Goto and Masakazu Ishihata and Shunsuke Kanda and Dominik K{\"o}ppl and Takaaki Nishimoto},
  title     = {Computing {NP}-Hard Repetitiveness Measures via {MAX-SAT}},
  booktitle = {30th Annual European Symposium on Algorithms (ESA 2022)},
  series    = {LIPIcs},
  volume    = {244},
  pages     = {12:1--12:16},
  publisher = {Schloss Dagstuhl -- Leibniz-Zentrum f{\"u}r Informatik},
  year      = {2022},
  doi       = {10.4230/LIPIcs.ESA.2022.12}
}

@article{ShibataB26,
  author       = {Hiroki Shibata and
                  Hideo Bannai},
  title        = {String Representation Based on Substring Equation Systems},
  journal      = {CoRR},
  volume       = {abs/2604.04377},
  year         = {2026},
  eprinttype   = {arXiv},
  eprint       = {2604.04377v5}
}

@article{GawrychowskiKRRW20,
  author    = {Pawe{\l} Gawrychowski and Tomasz Kociumaka and Jakub Radoszewski and Wojciech Rytter and Tomasz Wale{\'n}},
  title     = {Universal reconstruction of a string},
  journal   = {Theoretical Computer Science},
  volume    = {812},
  pages     = {174--186},
  year      = {2020},
  doi       = {10.1016/j.tcs.2019.10.027}
}

@article{Navarro22,
  author    = {Gonzalo Navarro},
  title     = {Indexing Highly Repetitive String Collections, Part {I}: Repetitiveness Measures},
  journal   = {ACM Computing Surveys},
  volume    = {54},
  number    = {2},
  pages     = {1--31},
  year      = {2022},
  doi       = {10.1145/3434399}
}

@article{BodlaenderCKN15,
  author       = {Hans L. Bodlaender and
                  Marek Cygan and
                  Stefan Kratsch and
                  Jesper Nederlof},
  title        = {Deterministic single exponential time algorithms for connectivity
                  problems parameterized by treewidth},
  journal      = {Inf. Comput.},
  volume       = {243},
  pages        = {86--111},
  year         = {2015},
  doi          = {10.1016/J.IC.2014.12.008},
  bibsource    = {dblp computer science bibliography, https://dblp.org}
}

@article{Schwartz80,
  author       = {Jacob T. Schwartz},
  title        = {Fast Probabilistic Algorithms for Verification of Polynomial Identities},
  journal      = {J. {ACM}},
  volume       = {27},
  number       = {4},
  pages        = {701--717},
  year         = {1980},
  url          = {https://doi.org/10.1145/322217.322225},
  doi          = {10.1145/322217.322225},
  bibsource    = {dblp computer science bibliography, https://dblp.org}
}

@article{CyganKW09,
  author       = {Marek Cygan and
                  Lukasz Kowalik and
                  Mateusz Wykurz},
  title        = {Exponential-time approximation of weighted set cover},
  journal      = {Inf. Process. Lett.},
  volume       = {109},
  number       = {16},
  pages        = {957--961},
  year         = {2009},
  doi          = {10.1016/J.IPL.2009.05.003},
  bibsource    = {dblp computer science bibliography, https://dblp.org}
}

@article{ZivL77,
  author       = {Jacob Ziv and
                  Abraham Lempel},
  title        = {A universal algorithm for sequential data compression},
  journal      = {{IEEE} Trans. Inf. Theory},
  volume       = {23},
  number       = {3},
  pages        = {337--343},
  year         = {1977},
  doi          = {10.1109/TIT.1977.1055714},
  bibsource    = {dblp computer science bibliography, https://dblp.org}
}

@article{Chen26,
  author       = {Xiangle Chen},
  title        = {Exact Algorithms and Complexity of the Smallest Grammar Problem},
  journal      = {CoRR},
  volume       = {abs/2610.32298},
  year         = {2026},
  eprinttype   = {arXiv},
  eprint       = {2610.32298}
}

@article{Williams05,
  author       = {Ryan Williams},
  title        = {A new algorithm for optimal 2-constraint satisfaction and its implications},
  journal      = {Theor. Comput. Sci.},
  volume       = {348},
  number       = {2-3},
  pages        = {357--365},
  year         = {2005},
  doi          = {10.1016/J.TCS.2005.09.023},
  bibsource    = {dblp computer science bibliography, https://dblp.org}
}

\clearpage

\appendix
\section{Details of the Exact Algorithms under Additional Constraints}
\label{sec:ses-fixed-dp}

We use the contracted graph $G_{\mathcal A}$ from \cref{sec:ses-additional} and assume $d \ge 2$.
We define the reduced incidence matrix $B_{\mathcal A} \in \mathbb Z^{[d-1] \times E(w)}$ of $G_{\mathcal A}$ by
\[
B_{\mathcal A}[u, e_i] = \begin{cases}
-1 & \text{if } u = p' \text{ and } p' < q',\\
1 & \text{if } u = q' \text{ and } p' < q',\\
0 & \text{otherwise},
\end{cases}
\]
where $u \in [d-1]$, $p' = \operatorname{idx}_{\mathcal A}(p)$, and $q' = \operatorname{idx}_{\mathcal A}(q)$ for $e_i = \{p, q\}$ with $p' \le q'$.

For $T = \{e'_{j_1}, \dots, e'_{j_\ell}\} \subseteq E_{\mathcal A}$ with $j_1 < \dots < j_\ell$ and $0 \le \ell \le d-1$, we define
\begin{align*}
v(T, \pi_1, \pi_2)
&= \left(\sgn(\pi_1)\prod_{h=1}^{\ell} B_{\mathcal A}[\pi_1(h), e_{j_h}]\right)
\left(\sgn(\pi_2)\prod_{h=1}^{\ell} B_{\mathcal A}[\pi_2(h), e_{j_h}]\right), \\
v_y(T, \pi)
&= \sgn(\pi)\prod_{h=1}^{\ell} B_{\mathcal A}[\pi(h), e_{j_h}]y_{e_{j_h}}
\end{align*}
for $\pi_1, \pi_2, \pi \in P_{d-1,\ell}$.
Both values are $1$ for the empty edge set and empty sequences.
These are the terms in the determinant expansions defining $N_k$ and $R_k(y)$.

For a nonempty edge set $T$, let $i$ be the index of its last edge and $k = |\mathcal Q(T)|$.
For $S_1, S_2, S \subseteq [d-1]$, we define $D[i, S_1, S_2, k]$ as the sum of $v(T, \pi_1, \pi_2)$ over the triples with these values of $i$ and $k$ satisfying $\Set(\pi_1) = S_1$ and $\Set(\pi_2) = S_2$.
We define $A[i, S, k]$ as the sum of $v_y(T, \pi)$ over the pairs with the same conditions on $i$ and $k$ satisfying $\Set(\pi) = S$.
The tables group the terms by the last edge and the sets of row indices used in the determinant expansions.

We set $D[0, \emptyset, \emptyset, 0] = A[0, \emptyset, 0] = 1$ and all other entries with $i = 0$ to $0$.
We also set entries with $k \notin [0, d-1]$ to $0$ and set $D[i, S_1, S_2, k] = 0$ when $|S_1| \ne |S_2|$.
These initial values correspond to the empty edge set and empty sequences.

For $i \in [|E(w)|]$, we compute $D[i, S_1, S_2, k]$ and $A[i, S, k]$ by

\begin{align}
D[i, S_1, S_2, k]
&= \sum_{j=0}^{i-1} \sum_{\substack{u \in S_1 \\ v \in S_2}}
(-1)^{\mathrm{rank}(S_1 \setminus \{u\}, u) + \mathrm{rank}(S_2 \setminus \{v\}, v)} \notag\\
&\qquad\cdot B_{\mathcal A}[u, e_i] B_{\mathcal A}[v, e_i]
D[j, S_1 \setminus \{u\}, S_2 \setminus \{v\}, k-c(j, i)],
\label{eq:ses-fixed-dp}\\
A[i, S, k]
&= y_{e_i}\sum_{j=0}^{i-1}\sum_{u \in S}
B_{\mathcal A}[u, e_i](-1)^{\mathrm{rank}(S \setminus \{u\}, u)} \notag\\
&\qquad\cdot A[j, S \setminus \{u\}, k-c(j, i)].
\label{eq:ses-fixed-rand-dp}
\end{align}
Here $j$ is the index of the last edge after removing $e'_i$ and $u, v$ are the last row indices in the two determinant terms.

\sesfixeddpcorrect*
\begin{proof}
We prove the recurrences by induction on $i$.
For $i = 0$, the empty edge set and empty sequences contribute $1$ with $k = 0$ and all other sums are empty.
This agrees with the initial values.

For $i \ge 1$, consider an edge set $T$ whose last edge is $e'_i$ and let $T' = T \setminus \{e'_i\}$.
Let $j$ be the index of the last edge of $T'$ with $j = 0$ when $T'$ is empty.
Each run occupies a consecutive interval in the edge order and hence $|\mathcal Q(T)| = |\mathcal Q(T')| + c(j, i)$.
Removing the last row indices $u, v$ from $\pi_1, \pi_2$ leaves the row sets $S_1 \setminus \{u\}$ and $S_2 \setminus \{v\}$.
As in \eqref{eq:ses-prefix}, the removed factors are $B_{\mathcal A}[u, e_i]B_{\mathcal A}[v, e_i]$ and the sign changes by $(-1)^{\mathrm{rank}(S_1 \setminus \{u\}, u)+\mathrm{rank}(S_2 \setminus \{v\}, v)}$.
Every triple contributing to $D[i, S_1, S_2, k]$ is obtained uniquely by adding these last entries to a shorter triple.
The induction hypothesis therefore gives \eqref{eq:ses-fixed-dp}.
For $A[i, S, k]$ the same argument removes one row index and the factor $y_{e_i}B_{\mathcal A}[u, e_i](-1)^{\mathrm{rank}(S \setminus \{u\}, u)}$.
This gives \eqref{eq:ses-fixed-rand-dp}.

For complete row sets the edge sets have size $d-1$ and the sequences are permutations of $[d-1]$.
The Leibniz formula gives $(\det B_{\mathcal A,T})^2$ after summing over the two permutations in $D$ and gives $\det B_{\mathcal A,T}\prod_{e'_i \in T}y_{e_i}$ after summing over the permutation in $A$.
Summing over the last edge includes each edge set $T$ exactly once and gives the stated identities for $N_k$ and $R_k(y)$.
\end{proof}

\end{document}